\documentclass{aa}

\usepackage{txfonts}
\usepackage{lipsum}
\usepackage{subcaption}         
\usepackage{amsmath}

\makeatletter
\DeclareRobustCommand{\HI}{%
  \mbox{H\check@mathfonts\fontsize\sf@size\z@\selectfont I} %
}
\makeatother                                
\usepackage{lscape}             
                                
\usepackage{placeins}

\usepackage[english]{}
\usepackage{natbib}
\usepackage[utf8]{inputenc}
\usepackage[colorlinks,urlcolor=magenta,citecolor=blue,linktocpage,breaklinks]{hyperref}

\usepackage{xspace}
\usepackage[]{footnote}
\usepackage{booktabs}
\usepackage{orcidlink}
\usepackage{float}
\usepackage{placeins}

\DeclareUnicodeCharacter{2212}{-}

\begin{document}

   \title{The explosive growth of the Messier 74 galaxy}

   \subtitle{A galaxy doubling its size in less than a Gigayear}

   \author{Ignacio Ruiz Cejudo \inst{1,2}\orcidlink{0009-0003-6502-7714}
   \and Ignacio Trujillo \inst{1,2}\orcidlink{0000-0001-8647-2874}
   \and Sergio Guerra Arencibia \inst{1,2}\orcidlink{0009-0001-7407-2491}
   \and Miguel R. Alarcon \inst{1,2,3}\orcidlink{0000-0002-8134-2592} 
   \and Miquel Serra-Ricart \inst{1,2,3}\orcidlink{0000-0002-2394-0711}
   \and Ouldouz Kaboud\inst{4}\orcidlink{0009-0007-7712-0683}
   }

   \institute{Instituto de Astrof\'{i}sica de Canarias, V\'{i}a L\'{a}ctea S/N, E-38205 La Laguna, Spain 
    \and
    Departamento de Astrof\'{i}sica, Universidad de La Laguna, E-38206 La Laguna, Spain 
    \and
    Light Bridges S.L., Observatorio del Teide, Carretera del Observatorio S/N, E-38500, Güímar, Tenerife, Canarias, Spain
    \and
    Department of Physics and Astronomy, University of Padova, Vicolo Osservatorio 3, I-35122 Padova, Italy
    }

   \date{Received March 11, 2026; Accepted June 16, 2026}

  \abstract
   {Galaxy formation models predict that galaxies grow inside-out, becoming larger over time. While observations broadly support this paradigm, the nature and timescales of this growth remain poorly constrained. We report the discovery of an extremely faint and young ($\sim$600 Myr) stellar component in the outer regions of the nearby galaxy Messier 74 (M74). Using deep optical imaging from the TST telescope at the Teide Observatory, reaching surface brightness limits of $\sim$30–31.5 mag arcsec$^{-2}$ in the \textit{g, r,} and \textit{i} bands, we detect stellar emission extending well beyond the previously known disc radius of $\sim$14 kpc. This newly identified component reaches galactocentric distances of $\sim$30 kpc, effectively doubling the known size of the stellar disc and matching the extent of the \HI disc. The revised size of M74 places it  in the upper envelope of the  mass–size relations. The young age of the outer stellar population suggests a recent episode of disc growth, potentially occurring on timescales shorter than $\sim$1 Gyr. We discuss a possible scenario in which a past flyby interaction with UGC~1176 may have triggered this extended star formation. Further studies of galaxies with similarly deep imaging will be key to determining whether such rapid outer disc growth is common or exceptional.}

   \keywords{galaxies: evolution - galaxies: star formation - galaxies: stellar content - galaxies: structure - galaxies: individual: M74 - techniques: image processing}

   \maketitle
\nolinenumbers

 \section{Introduction}\label{sec:int}

Galaxy formation models predict that galaxies should become larger as cosmological time progresses. \citet{1980Fall} suggested that galaxies form when low angular momentum gas collapses into a dense central component. As the Universe evolves, subsequent gas accretion occurs, with progressively higher specific angular momentum. This gas settles into extended orbits and fuels star formation in the galactic outskirts \citep[see, e.g., ][]{1998Mo}. This process results in a continuous increase in the disc scale length. This theoretical expectation has been supported by observational evidence. Studies such as \citet{2006ApJ...650...18T}, \citet{2008ApJ...687L..61B},  \citet{vdw14} or \citet{lange15} showed a mild increase in effective radius for Milky Way-like galaxies since z$\sim$1. Alternatively, \citet{2024Buitrago} using the location of the edge of the star formation threshold as a proxy for size, observed a strong increase in disc sizes from z$\sim$1 to the local Universe (around a factor of two for Milky Way-like galaxies).  \\

The nature of this growth is still a matter of debate. Is this a gentle process where disc grows slowly with time? Or on the contrary, the growth is very fast and episodic? In the absence of external perturbations, gas accretion is directly connected to the initial conditions during collapse and cooling \citep{1998Mo}. Consequently, the disc grows inside-out through continuous star formation from the accreted gas \citep{2018genel,2013Bird,2011Pichon}. However, discs are susceptible to mergers. Late-type galaxies can accrete cold gas from gas-rich satellites, fuelling star formation in a larger disc \citep{2006roberson,2008Sancisi}. If this is the case, young stellar populations would be expected in the outskirts of late-type galaxies. One of the observational signatures of these events has been suggested to be the extended ultraviolet (XUV) discs \citep{gildepaz+05,thilker+05}, which exhibit UV emission beyond the optical extent of the galaxy (defined by the $D_{25}$ isophote). \citet{thilker07} reported that $\gtrsim20\%$ of their sample of late-type galaxies host an XUV disk. They also linked this emission to recently formed stellar populations fuelled by minor mergers.    \\

The definition of the XUV discs is, however, inherently tied to the definition of the optical size. The classic proxies used for measuring the size of galaxies, such as the aforementioned $R_{25}$ \citep{1936redman} or the effective radius \citep[$R_{e}$, ][]{1948devaucoleurs}, are strongly dependent on the observed wavelength \citep[see, e.g., ][]{2015kennedy}. To overcome this problem, \citet{2020trujillo} proposed a new definition for the boundary of a galaxy, based on the limits (present or past) of \textit{in situ} star formation. If a sudden decrease in the surface density of cold gas is present, then the \textit{in situ} star formation will experience a rapid drop \citep[see, e.g., ][]{2008Roskar}. Consequently, \citet{2020trujillo} used the star formation density threshold \citep[$\Sigma_{\HI}\sim3-10\,M_{\odot}\rm{pc^{-2}}$, ][]{2004Schaye} to define a proxy for the edge of the galaxy. They showed that this new size definition ($R_{1}\equiv R(\Sigma_{*}=1\, M_{\odot}\,\rm{pc}^{-2})$, assuming a star formation efficiency of 0.3) reduced the scatter in the mass-size relation. Building on this, \citet{chamba22} used the location of the outermost truncation in the surface brightness profiles, $R_{edge}$, as a new proxy for galaxy size. Their work also noted that location of $R_{edge}$ corresponds roughly to $R_{1}$ for late-type galaxies with stellar masses close to the Milky Way, suggesting its connection to the star formation threshold in star-forming disks.  \\

Detecting such low stellar mass densities requires deep imaging to achieve surface brightness depths fainter than $\mu_{g}$$\sim$26-27\,$\rm{mag\,arcsec^{-2}}$ \citep[see, e.g., ][]{2025Golini,2025Raji}. In this work, we present a deep optical study of the faint stellar component in the outskirts of Messier 74 (M74, NGC~628). M74 is a grand design spiral galaxy \citep{1994cag..book.....S} of type SA(s)c. It is located at a distance of $9.3\pm1.8$ Mpc \citep[from type II supernovae, ][]{2005Hendry} and has an estimated stellar mass of $1.5-2.2\times10^{10}\,M_{\odot}$ \citep{2015Querejeta,2019Leroy}. There are several indications that this galaxy is experiencing a major stellar population build-up. In the last two decades, three type II supernovae (SN 2003gd, 2013ej and 2019krl) have been observed in M74, as well as one type Ic \citep[SN 2002ap, see ][]{2020Michalowski}. It is also classified as a type I XUV disc by \citet{thilker07}. Various observations have revealed the complexity of its \HI component, featuring a warp that may be signature of recent gas accretion \citep{kamphuis92,2006Auld}. We aim to study the faint stellar component in the outskirts of the disc using ultra-deep optical imaging. \\

The paper is structured as follows. In Sec.~\ref{sec:data} we present the optical imaging of M74, as well as the \HI and UV data used. In Sec.~\ref{sec:cirrus} we present the application of a methodology to mitigate the effects of Galactic Cirrus. In Sec.~\ref{sec:res} we show the surface brightness, colour and stellar surface density radial profiles derived from the optical and UV data. In Sec.~\ref{sec:discusion} we analyse the results in the outer regions of M74, and in Sec.~\ref{sec:conclusions} we summarize our main conclusions. Unless explicitly stated, all magnitudes are given in the AB system \citep{1983Oke}.

\section{Data}\label{sec:data}

In this work, we present new deep optical imaging data obtained with the commissioning Transient Survey Telescope (TST)\footnote{\url{https://tst.iac.es}}, a 1-m, f/1.3
Ritchey-Chrétien telescope located at the Teide Observatory,
Canary Islands (Lat. $28^\circ 18^{\prime} 04^{\prime\prime}$ N, Long. $16^\circ 30^{\prime} 38^{\prime\prime}$W). Observations were conducted using the prime focus camera FERVOR-L (\textit{Fast Embedded-sCMOS Robotic Visible Observatory for Rapid transients}), based on sCMOS sensor \citep[Sony IMX411, $14303\times10748$ pixels, ][]{Alarcon_2023}. FERVOR-L has a pixel scale of 0.60 arcsec~pix$^{-1}$ and a field of view (FoV) of  $2\fdg4\times1\fdg8$. Due to the large size of the images, we applied binning to the data to make the data reduction feasible for the available computational resources. We adopted a binning factor of 3 (i.e, a pixel scale of $1.8\,\rm{arcsec\,pix^{-1}}$). This factor reduces the resources needed while preserving the resolution of the structures studied in this work.  \\

M74 was observed between August 20$^{\rm{th}}$ and $29^{\rm{th}}$, 2025, using standard Sloan $g^{\prime}$, $r^{\prime}$ and $i^{\prime}$ filters (three nights per band) manufactured by Baader Planetarium (hereafter $g$, $r$ and $i$). The total exposure times for the stacked images are 6.4~h ($g$ band), 5.4~h ($r$ band) and 6.5~h ($i$ band). The $3\sigma$ surface brightness limiting magnitudes reached in the stacked images are 31.44, 30.58 and 29.85 mag arcsec$^{-2}$ in the $g$, $r$ and $i$ bands respectively (in regions equivalent to an area of $1\arcmin\times1\arcmin$). A summary of exposure, depths and seeing of the stacked images can be found in Table~\ref{tab:summary}.\\

\begin{table}
    \caption{Summary of M74 observations with TST.}
    \label{tab:summary}
    \centering
    \begin{tabular}{cccc}
    \hline
    \hline
    Filter & $t_{exp}$ & $\mu_{lim}\,(3\sigma,1\arcmin\times1\arcmin)$ & FWHM \\
     & [h] & [mag arcsec$^{-2}$] & [arcsec] \\
     \hline
     \textit{g} & 6.4 & 31.44 & 2.3 \\
     \textit{r} & 5.4 & 30.58 & 2.3 \\
     \textit{i} & 6.5 & 29.85 & 2.2 \\
     \hline   
    \end{tabular}
\end{table}

Since this is one of the first works utilizing TST data, we give a detailed description of the observation strategy and data reduction in the following sections.

\subsection{Observation strategy} \label{subsec:obstrat}
The observing strategy used to achieve sensitivity to low surface brightness (LSB) detections follows the prescriptions outlined by \citet{2016trujillo}. This strategy is designed both to maximize the time  spent observing M74 and to accurately characterize the sky background during data collection. Part of this characterization consists of using the science images for the flat field correction (see Sec.~\ref{subsubsec:flatbias}). To ensure that every pixel on the detector is free of sources in at least some exposures, we applied a dithering pattern with offsets up to $0\fdg5$, which is larger than the apparent size of M74 \citep[$D_{25}\sim0\fdg2$, ][]{1973Nilson}. The exposure time per image was set to 1 minute, which minimizes saturation and helps in characterizing the sky at the time of observation.

\subsection{Data reduction}\label{subsec:datared}
The data was reduced following a LSB friendly pipeline, in order to preserve the faint structures of the field and minimize the contamination. The main steps are described in the following subsections. 

\subsubsection{Bias and flat-field correction}\label{subsubsec:flatbias}
The first step was to handle saturated pixels and perform the bias correction. All pixels with values above 585000 (this is equivalent to combine nine saturated original pixels into a one binned pixel, i.e., $\simeq9\times2^{16}$) Analog-to-Digital Units (ADU) were first masked. We constructed a master bias and subtracted it from the science frames. \\

Following the strategies of other LSB reductions \citep[see, e.g, ][]{2021trujillo,2025junais}, we performed the flat-field correction using the science images themselves. First, we normalized the science images using the resistant mean in a region of constant illumination (a ring with $R=1100$ pixels and width of 200 pixels). Then, the normalised, bias-corrected images were combined to build the flat field. We followed the \textit{Running Flat} strategy described in \citet{2025junais} and \citet{2025Saremi}: we used the \textit{N} previous and following exposures of each frame to build its corresponding flat. This number depended on the band: 10, 8 and 6 for the $g$, $r$ and $i$ bands, respectively. \\

The flat fielding correction was iterated three times. In the first iteration, images were stacked using a sigma-clipped median. In the second iteration, we masked all detected sources in the flat-corrected images from the first iteration prior to the stacking. The masks were generated using \texttt{Gnuastro}'s \texttt{NoiseChisel} \citep{gnuastro,noisechisel}. The masked images were then stacked using a sigma-clipped median to build the flat field. Since each iteration produces a better flat field, the quality of the data is enhanced, yielding a better mask. Consequently, we performed a third iteration building the masks from the images corrected with the second iteration flats. Finally, to address vignetting issues, all pixels where the flat-field illumination was below $70\%$ were masked.

\subsubsection{Astrometry}\label{subsubsec:astrometry}
The data produced by the TST telescope is already astrometrically calibrated. However, to improve the relative astrometry between images as much as possible, we computed the geometric distortion of the camera using \texttt{SCAMP} \citep[v2.10.0; ][]{scamp}. \texttt{SCAMP} reads the catalogs generated by \texttt{SExtractor} \citep[v2.25.0; ][]{sextractor} and computes the SIP coefficients for the astrometric solution. This process was repeated three times, with each iteration improving the previous solution. Finally, we used \texttt{SWarp} \citep[v2.41.5; ][]{swarp} to resample the images onto a common grid of $7001\times7001$ pixels, fixing the pixel scale to $1.794\,\rm{arcsec\,pix}^{-1}$. The resampling method selected was LANCZOS3.

\subsubsection{Sky Subtraction and Coaddition}\label{subsubsec:skysubco}
Sky subtraction is a crucial step in data reduction for LSB astronomy. If not performed carefully, this process can introduce artifacts and lead to oversubtraction, both of which can significantly impact the detection of LSB features in the field. We performed sky subtraction on each frame individually, under the assumption of a constant background. This assumption is reasonable, despite the large FoV of the camera, due to the steps taken in Sec.~\ref{subsubsec:flatbias} to correct the illumination of the images. To characterize the background, an aggressive mask was created for each individual frame. Then, we computed the sigma-clipped median of the unmasked pixels and subtracted it from the individual frames. \\

After astrometric calibration and sky subtraction, the frames were combined into a single coadd. We built the coadded image using the weighted mean of the individual frames, where the weight of the $i^{th}$ frame is $\sigma_{min}^{2}/\sigma_{i}^{2}$, with $\sigma_{i}$ representing the standard deviation of the background of each frame, and $\sigma_{min}$ being the minimum value among all the frames. This weighted mean combination is optimal for images with Gaussian-like noise \citep[see ][ Chap. 4.4.5]{leo1988techniques}. During this process, a sigma-clipping mask was applied to mask artifacts such satellite trails.\\

The stacked image is deeper than the individual exposures, increasing the signal-to-noise ratio of the image and revealing faint, previously undetected signal such as Galactic Cirrus. During the initial sky subtraction step, this signal was treated as background, leading to oversubtraction. To address this, we applied an iterative process to the sky subtraction and coaddition steps. After the first iteration, we used the stacked image to build a better mask, which was then applied to the individual frames prior to sky subtraction. Then, a new stacked image with an improved background was built and compared with the previous iteration. If the percentage of light recovered in this new iteration is lower than $\sim1\%$, we stop the iterative process and use the stacked image from the last iteration as our final coadd. Since the camera is more sensitive in the $g$ band, and Galactic Cirrus has a higher signal-to-noise ratio in this band, the iterative process was only applied to the $g$ band reduction. The final mask built from this reduction was then combined with the first-iteration masks for each of the other bands on a final iteration.

\subsubsection{Photometry calibration}\label{subsubsec:photocal}
The final step of the data reduction was to calibrate the images, computing the conversion factor from pixel values (ADUs) to physical units (Janskys). In this work, we accomplished this by comparing stellar fluxes in the individual images with their corresponding fluxes derived from the spectra in Gaia DR3 \citep{GaiaDR3}.\\

First, the available Gaia spectra of stars in the field were downloaded. These spectra were then convolved with the transmission curve of the respective filter to derive the expected magnitude for each star. Next, we measured the flux of the stars in the individual frames by applying aperture photometry. Following a strategy similar to that outlined in \citet{2025junais}, a large aperture was used to enclose the entire star. In this work, we selected apertures with a radius of $r=7R_{e}$. To use non-saturated stars with high SNR, we selected sources within a specific magnitude range (14.0 < mag < 15.5 for the $g$ and $r$ bands, and 14.0 < mag < 16.5 for the $i$ band). We then compared the measured flux of these stars with the magnitude obtained from the spectra to compute the calibration factor for each individual frame. Finally, we adopted a common calibration factor for all the individual frames, derived from the sigma-clipped mean of the individual calibration factors. The use of a common calibration factor is done by the assumption that the physical conditions of the camera are stable with time. Since data for each filter was obtained in consecutive nights and with the same instrumental set-up, this assumption is reasonable. The zero-point was set to 22.5 in the AB magnitude system. {We present the colour (\textit{g, r, i}) image of the coadded images in Fig.~\ref{fig:tstfield}.
\begin{figure*}
    \centering
    \includegraphics[width=\textwidth]{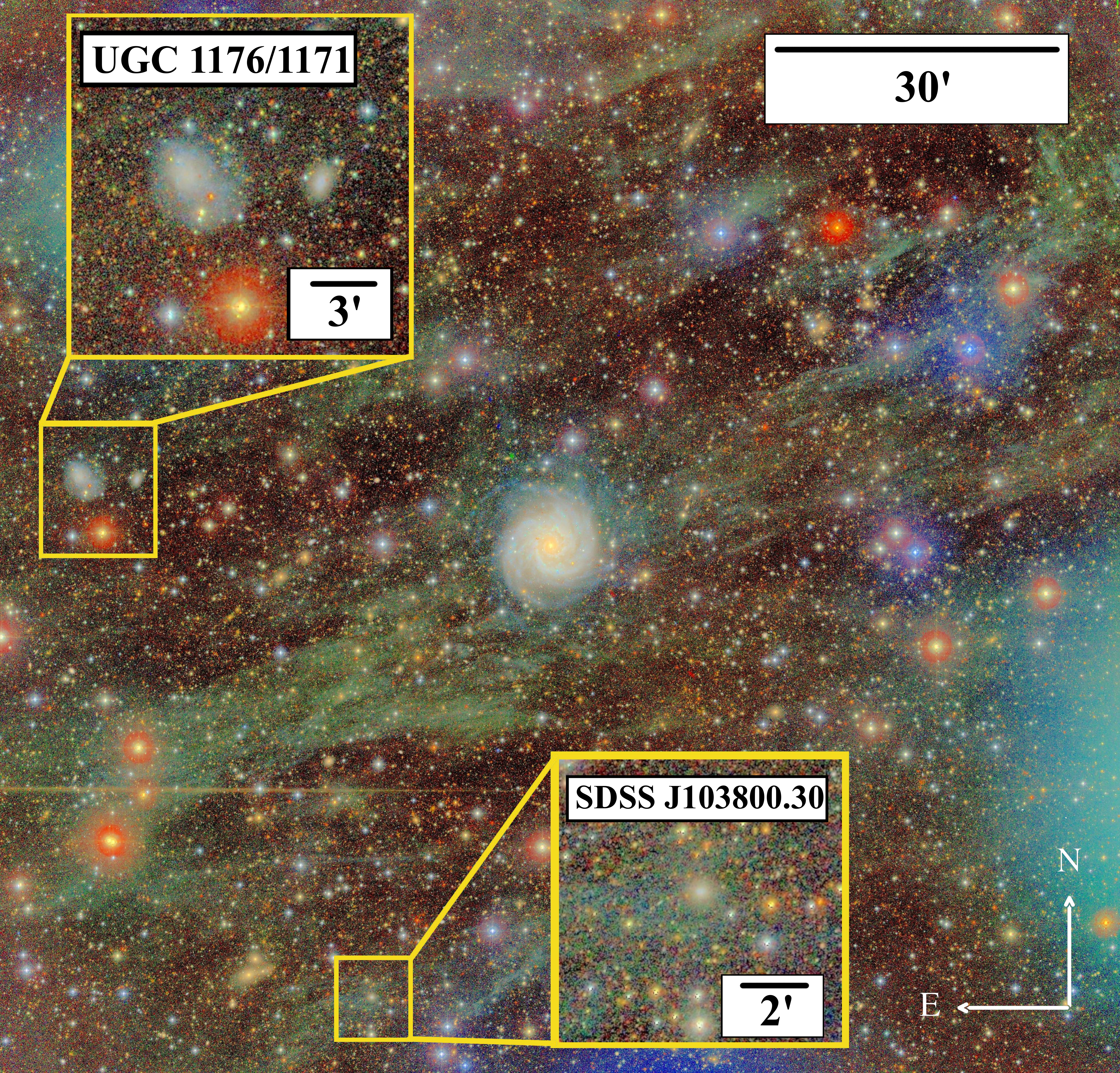}
    \caption{Colour (\textit{g, r, i}) image of M74 and its surroundings. The image was generated using \texttt{Gnuastro}'s script \texttt{astscript-color-faint-gray} \citep{astscript-color-faint-gray}. Insets show zoomed-in views of the UGC~1176/1171 pair (top left) and the galaxy SDSS~J013800.30+145858.1 (bottom), all of which are reported to belong to the M74 group \citep[see, e.g., ][]{2020Michalowski}}
    \label{fig:tstfield}
\end{figure*}
\subsection{Archival data: \HI and UV} \label{subsec:archdata}

We used archival \HI and Ultraviolet  data of the galaxy to complement the optical analysis. We retrieved the \HI map of M74 from THINGS \citep{things} survey. We solely used the integrated \HI map, with units of $M_{\odot}\,\rm{pc}^{-2}$. \\

The UV data were obtained from the GALEX satellite GR7 data release \citep{bianchi14}. We used images from the Guest Investigator Program GI3$\_$050001$\_$NGC628, with a total exposure time of $\sim$3.3 hours per band. The images were reprocessed using the methodology described in \citet{UVLIGHTS}, reaching $3\sigma$ depths of 31.5 and 31.4 mag arcsec$^{-2}$ for the FUV and NUV bands, respectively, in equivalent areas of $1\arcmin\times1\arcmin$. 

\section{Galactic Cirrus removal}\label{sec:cirrus}

Due to its position relatively close to the Milky Way plane ($b=-45.7^{\circ}$), the field of M74 is surrounded by Galactic Cirrus (see Fig.~\ref{fig:tstfield}). These Cirrus difficult the analysis of the low surface brightness features of M74 studied in this work. Because of that, we apply a novel technique developed by \citet{kaboud26} for the LIGHTS survey \citep{2021trujillo,Zaritsky2024}. This technique mitigates the effect of Galactic Cirrus by assuming a linear relation between optical and far-infrared emission of Cirrus, building on the approach by \citet{Mihos2017}. The far-infrared data used for subtracting the Cirrus are the Herschel's 250 $\mu$m maps \citep{Griffin2010,Boselli2010}, with a pixel scale of $6.0$ arcsec pix$^{-1}$ and an angular resolution of $18.1$ arcsec. \\

The methodology works as follows. First, we re-scaled the optical data into the Herschel resolution ($6.0$ arcsec pix$^{-1}$) and convolved them with a Gaussian kernel with a FWHM of $18.1$ arcsec, in order to account for the PSF of the Herschel data. Then, we isolated the Cirrus emission in the optical and in the far-infrared image by masking the rest of light sources. The masks are built individually for TST and Herschel data, and combined into a single mask. We used the survival pixels to find the relation $F_{op}=\alpha F_{250}$, with both fluxes in units of MJy/sr. We find factors of $\alpha=4.5,\,9.0,\,9.8\times10^{-4}$ for $g$, $r$ and $i$ bands respectively. Finally, we multiplied Herschel data by these factors and subtracted the map from the optical data only re-scaled to the Herschel resolution. However, in addition to dust emission, the Herschel $250\,\mu\rm{m}$ contains flux from high-redshift background objects and the central galaxy itself. Therefore, we need to fill in those locations with only dust emission. For compact objects, this is done by interpolating the flux of surrounding dust pixels. In the case of the central galaxy, we compute the mean flux of dust in a ring surrounding the galaxy, and fill the pixels with this value.  \\

It is important to note that this process decreases the spatial resolution of our images. In this study, the objects of interest are large enough ($\sim\,\text{arcmin}^{2}$) to not be affected in their analysis by this resolution decrease. A summary of the process in images can be found in Fig.~\ref{fig:cirrusrem}. A comparison between the profiles presented in Sec.~\ref{sec:res} and the ones with Galactic Cirrus is included in Appendix ~\ref{ap:cirrus}.

\begin{figure*}
    \centering
    \includegraphics[width=\textwidth]{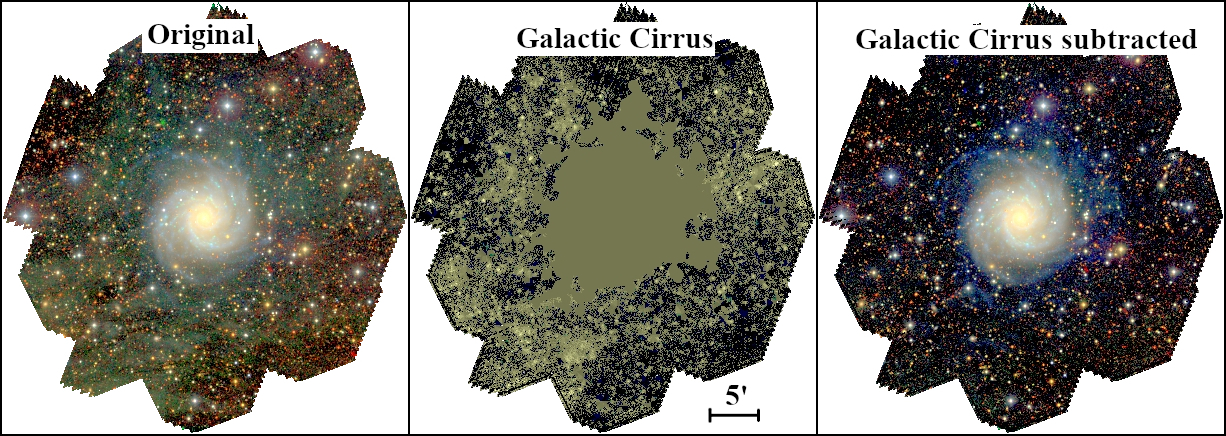}
    \caption{Modeling and subtraction of the Galactic Cirrus in M74 field. The left panel shows the optical image at $6\arcsec$. The central panel shows the Herschel 250 $\mu\text{m}$ image, with the expected Galactic Cirrus contribution on top of the compact sources and on top of the central galaxy adjusted as explained in the text. The right panel shows the "dust-corrected" optical image after subtracting the central image from the one on the left.}
    \label{fig:cirrusrem}
\end{figure*}
\section{Results}\label{sec:res}

We present the deep optical image of the M74 field in Fig.~\ref{fig:tstfield}. We observe previously reported satellites \citep{2020Michalowski}. The UGC~1176/1171 pair is located in the eastern part of the field, while SDSS~J013800.30+145858.1 is located to the south of M74. The field is also permeated by Galactic Cirrus, seen as green plumes, with a mean surface brightness of $\mu_{g}\sim27\,\rm{mag\, arcsec^{-2}}$ and a mean colour $g-r\sim0.7$ in the brightest parts. After subtracting them with the techniques explained in Sec.~\ref{sec:cirrus}, two different components appear of M74. The inner region, enclosed by a red circle in Fig.~\ref{fig:m74}, corresponds to a classic disc with blue spiral arms. Surrounding this classic disc, new faint bluer regions appear (enclosed by the red and yellow circles in Fig.~\ref{fig:m74}). These regions have the shape of spiral arms in the east side of M74, while on the west part the structure is more irregular. \\

\begin{figure}
    \centering
    \includegraphics[width=\linewidth]{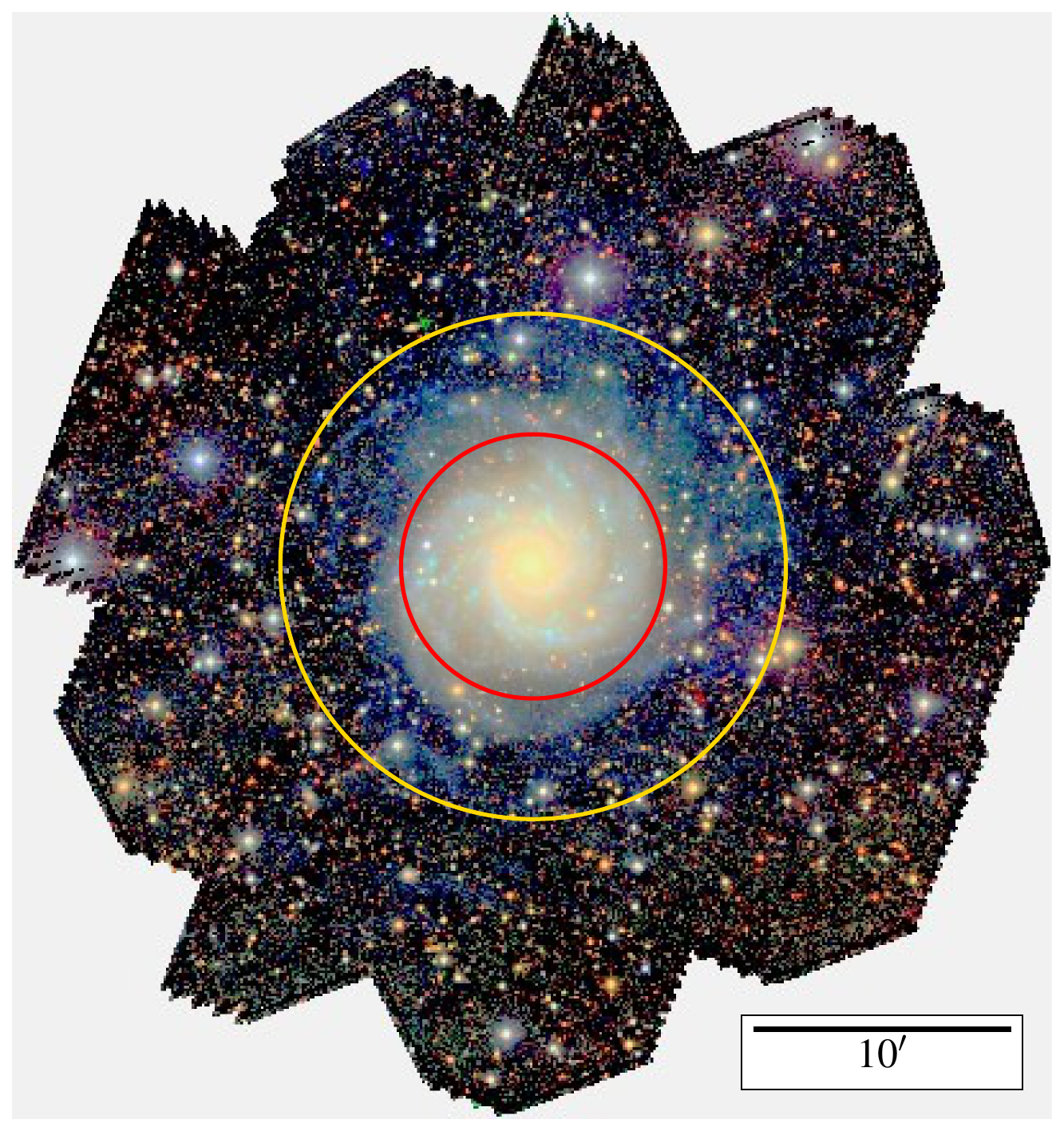}
    \caption{$gri$ colour composite image of M74 after Galactic Cirrus removal. The red circle corresponds to the radius at which the surface brightness profile in $g$ reaches 26 mag arcsec$^{-2}$ (see Fig.~\ref{fig:radprofs}, central panel). The yellow circle corresponds to the radius at which it reaches 29 mag arcsec$^{-2}$.}
    \label{fig:m74}
\end{figure}

In Fig.~\ref{fig:radprofs}, we present the surface brightness profiles of M74 in the $g$, $r$, $NUV$ and $FUV$ bands. The profiles are extracted in two different regions. The ones on the middle panel are extracted in circular apertures, whereas the ones in the right panel are extracted in a wedge-shaped sector of $20^{\circ}$ at an angle of $30^{\circ}$ (from west, anti-clockwise). The selection of this wedge is based on a visual inspection of Fig.~\ref{fig:m74}. In that direction, the blue region appear more homogeneous and bright, and the transition with the inner disc is more noticeable. Thus, this wedge-shaped sector appears as the most optimal one to identify the limits of the extension of these blue regions. The profiles are limited to values brighter than the limiting $3\sigma$ surface brightness limit corresponding to the area over which the profile was computed. \\  

\begin{figure*}
    \centering
    \includegraphics[width=\textwidth]{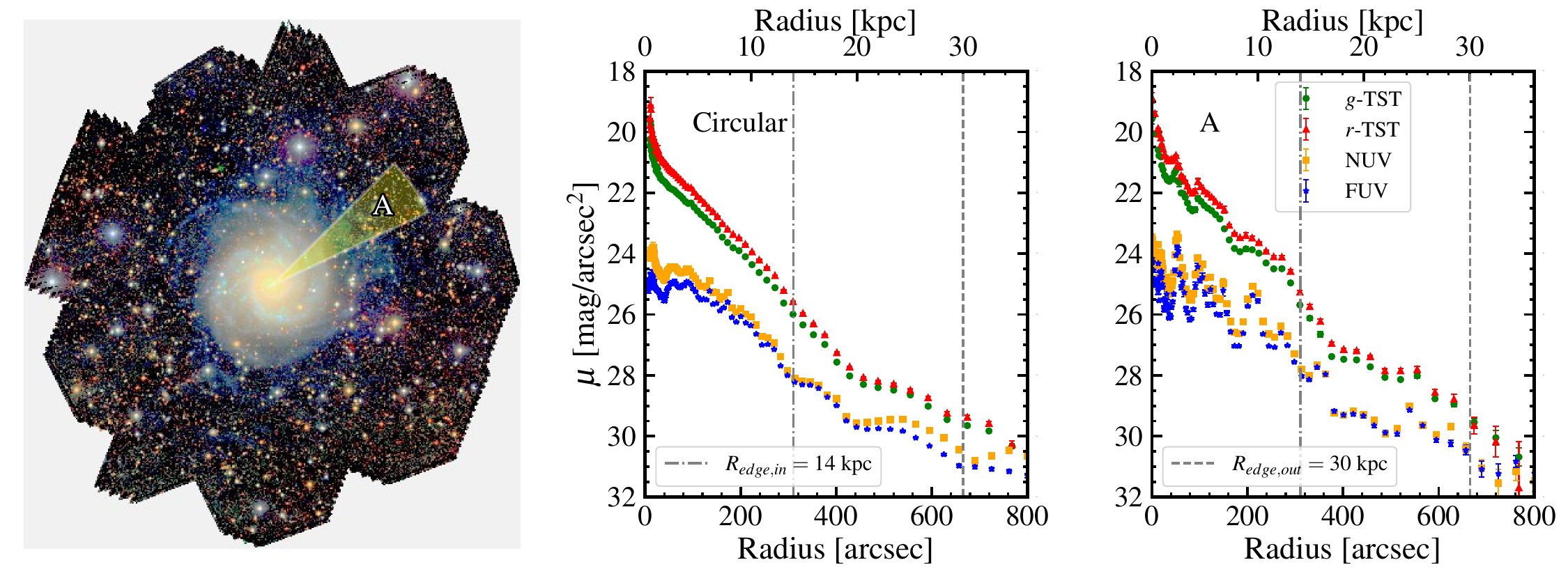}
    \caption{\textit{Left: }TST colour image of M74 showing the different sectors used to compute the surface brightness radial profiles. \textit{Middle and right: }$g$, $r$, $NUV$ and $FUV$ surface brightness radial profiles of M74 in circular apertures and along a wedge-shaped sector. Vertical lines indicate the position of the inner and outer discs discussed in Sec.~\ref{sec:res}.}
    \label{fig:radprofs}
\end{figure*}

 In Fig.~\ref{fig:colors}, we present the $(g-r)_{0}$ and $(FUV-NUV)_{0}$ radial profiles. These have been corrected for Galactic extinction using the \citet{cardelli+89} extinction curves for the Milky Way. For the UV bands, we adopted the corrections from \citet{bianchi2017}, i.e., $A_{FUV}=8.06E(B-V)$ and $A_{NUV}=7.95E(B-V)$. For the optical bands, we combined the \citet{cardelli+89} Milky Way extinction curves with the $g$ and $r$ filter responses to derive the reddening corrections. Using a reddening $E(B-V)=0.07$ (retrieved from the NASA/IPAC Extragalactic Database, NED), we obtained $A_{FUV}=0.564,\,A_{NUV}=0.557,\,A_{g}=0.253,\,\text{and}\,A_{r}=0.187$. \\

\begin{figure*}
    \sidecaption
    \includegraphics[width=12cm]{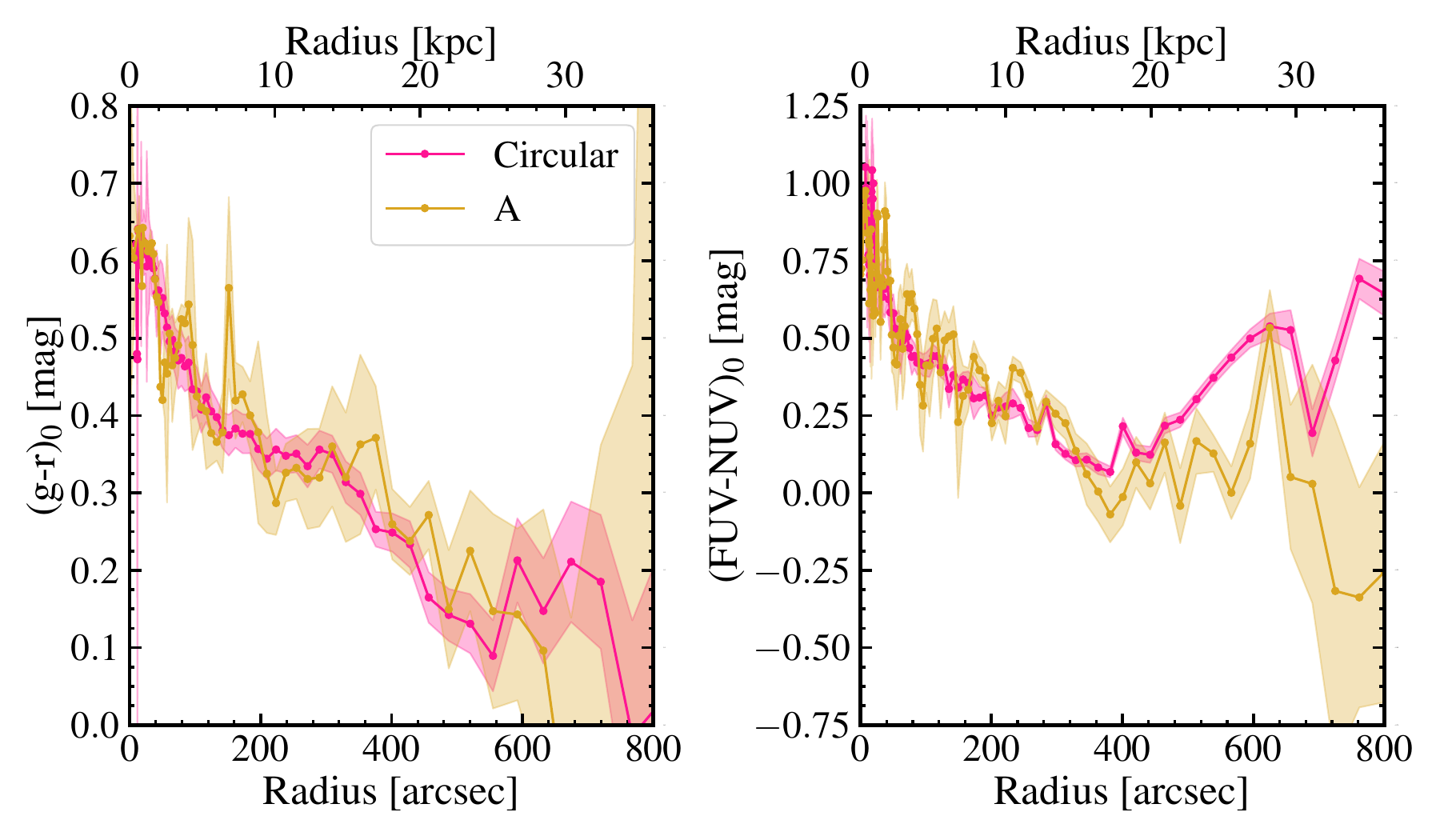}
    \caption{Radial colour profiles of M74. The left panel show the $(g-r)_{0}$ profiles, and the right panel display the $(FUV-NUV)_{0}$ profiles. }
    \label{fig:colors}
\end{figure*}

Finally, we compared the stellar surface density obtained from the optical data with the \HI surface density derived from the THINGS data. First, we calculated a mean stellar mass-to-light ratio $(M/L)$. Once this was determined, we computed the stellar surface density profile (in $\rm{M}_{\odot}\,\rm{pc}^{-2}$) using the following equation \citep[see e.g. ][]{Bakos08}:
\begin{equation}\label{eq:steldens}
\log\Sigma_{*}=\log(M/L)_{\lambda}-0.4\left(\mu_{\lambda}-\mu_{abs,\odot,\lambda}\right)+8.629\, ,
\end{equation}
where $\mu_{abs,\odot,\lambda}$ is the absolute magnitude of the Sun at wavelength $\lambda$.  
To compute $M/L$, we followed the procedure described by \citet{RoediguerCourteau2015}, using as the basis for our estimation the extinction-corrected $(g-r)_0$ profile and the extinction-corrected surface brightness profile in the $g$ band. We used the parameters provided by \citet{RoediguerCourteau2015} for the \citet[BC03]{bc03} models and a Chabrier initial mass function \citep{chabrier03}. Given that M74 is a nearly face-on galaxy, we did not apply any inclination correction. In Fig.~\ref{fig:steldens}, we compare the stellar mass surface density profiles with the \HI surface density profile. The uncertainties of the stellar mass surface density profiles are obtained from the propagation of the uncertainties of the $g$ band surface brightness profile and the $(g-r)_{0}$ colour profile (used for computing the mass-to-light ratio). Then, we add a $0.1$ dex error in the mass-to-light ratio, associated with the uncertainty of the mass-to-light versus colour relation discussed in \citet{RoediguerCourteau2015}.  \\

\begin{figure*}
    \sidecaption
    \includegraphics[width=12cm]{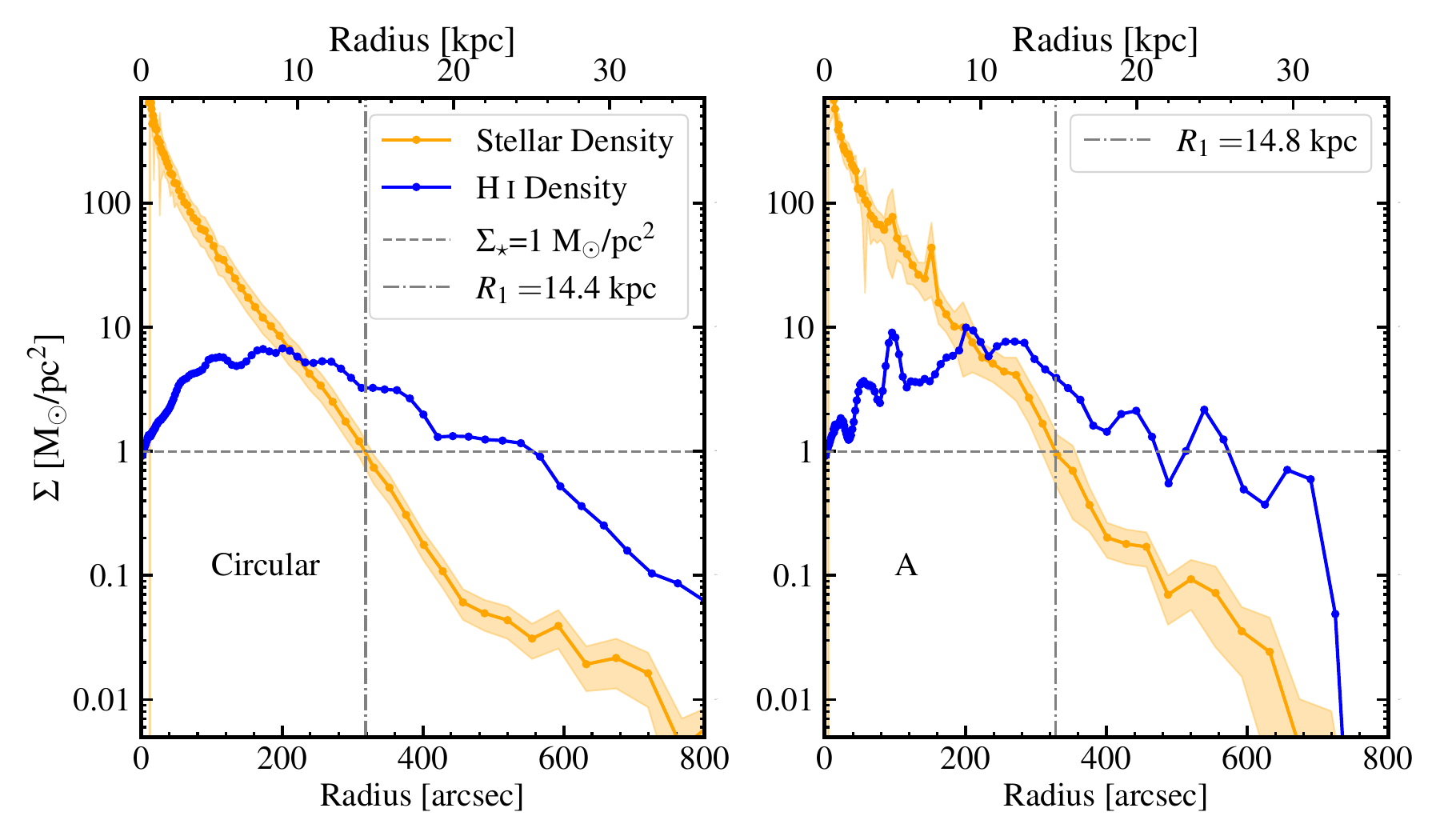}
    \caption{Stellar (yellow) and \HI (blue) surface density profiles of M74. A vertical and an horizontal line are included indicating the position of $R_{1}\equiv R(\Sigma_{*}=1\,M_{\odot}/\rm{pc}^{-2})$.}
    \label{fig:steldens}
\end{figure*}

The surface brightness (Fig.~\ref{fig:radprofs}) and stellar surface density (Fig.~\ref{fig:steldens}) profiles reveal two different behaviours. The circular apertures optical profiles decay exponentially up to $R\sim18\,\rm{kpc}$, while the UV profiles show two truncations at $R\sim12\,\rm{kpc}$ and $R\sim18\,\rm{kpc}$. In the A wedge, the optical profiles present a truncation at $R\sim14\,\rm{kpc}$, which we indicate as a vertical dashed-dotted line in both surface brightness profiles. Then, at $R\sim18\,\rm{kpc}$, all profiles present an anti-truncation followed by an almost flat behaviour. Finally, a break at $R\sim30\,\rm{kpc}$ is observed in the UV profiles and the optical profiles of the A sector, which we mark as a vertical dashed line in the surface brightness profiles in Fig.~\ref{fig:radprofs}. This break coincides with the visual extension of the blue component observed in the images (see Sec.~\ref{disc:size}).  \\

The colour profiles, in contrast, do not display these truncations as clearly. The general trend show a decay to bluer colours up to $R\sim14\,\rm{kpc}$. Then, the trend becomes more pronounced towards further radii. Uncertainties in the profiles prevent us from observing any reddening beyond the $R=30\,\rm{kpc}$ discussed on the previous paragraph.

\section{Discussion}\label{sec:discusion}

\subsection{Comparison with previous data}\label{disc:sdsscomp}
As shown in Fig.~\ref{fig:radprofs}, the outermost blue disk has a very low surface brightness ($\mu_{g}\sim28\,\rm{mag\,arcsec^{-2}}$). Thus, a LSB friendly observation strategy and reduction pipeline is key to preserve these faint structures. To enhance this, we compare our deep data with those from the \textit{Sloan Digital Sky Survey} \citep[SDSS; ] [DR16]{sdssDR16} in Fig.~\ref{fig:sdssvstst}. The data from SDSS is clearly shallower, $\mu_{g,lim}=29.5\,\rm{mag\,arcsec^{-2}}\,(3\sigma;1\arcmin\times1\arcmin)$, with artifacts from the reduction process present (such as the red gradient at the right size or the greenish background). In the SDSS, only the inner disc is visible, whereas the blue structure surrounding the galaxy is not perceptible.  \\

\begin{figure*}
    \centering
    \includegraphics[width=\textwidth]{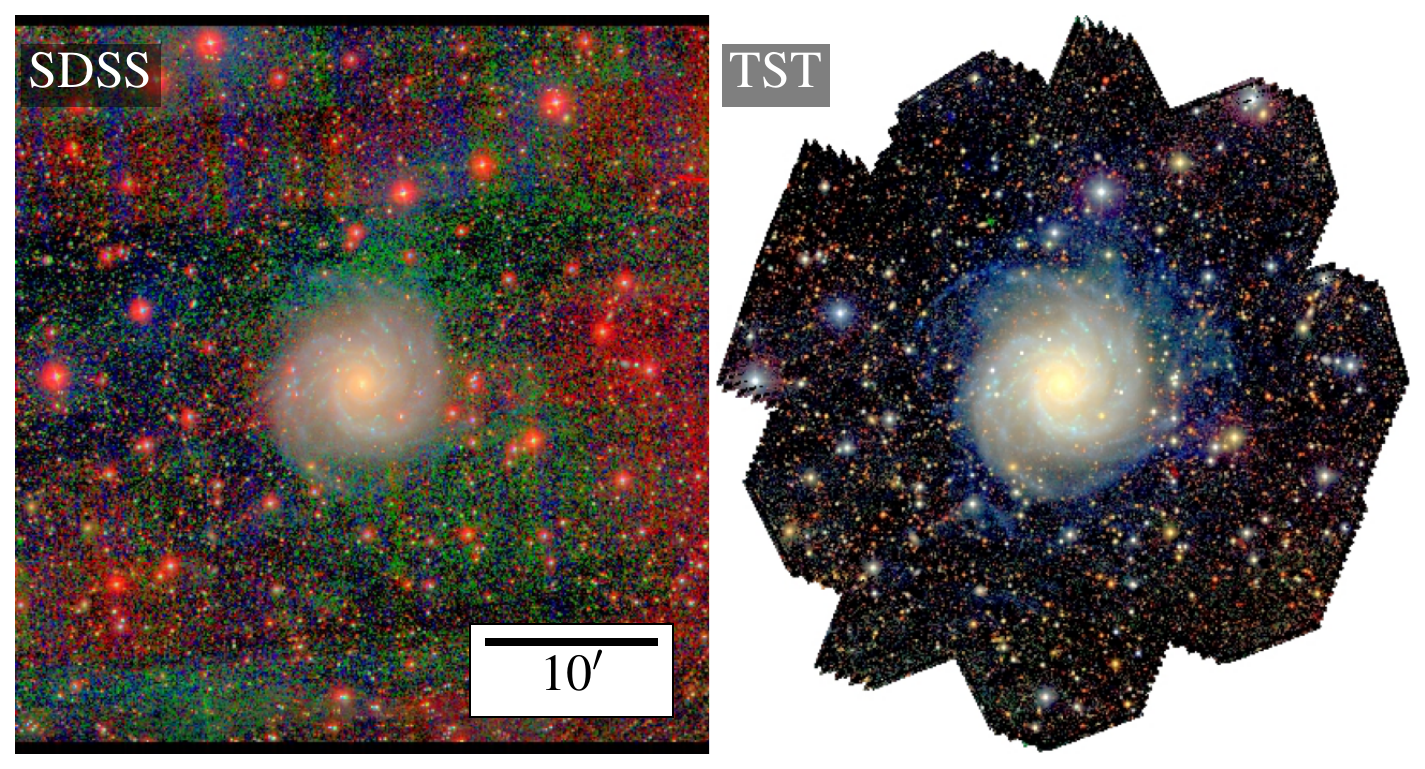}
    \caption{SDSS (left) and TST (right) colour composite (\textit{g, r, i}) images of M74. A 10-arcminute scale bar is provided for reference. In the TST image, the faint structures surrounding the galaxy (i.e., the newly discovered blue disc) are clearly visible.}
    \label{fig:sdssvstst}
\end{figure*}

We also compare the surface brightness, colour and stellar surface density profiles of our dataset with the SDSS data in Fig.~\ref{fig:sdssvststprofs}. To make a fair comparison, we use the TST images prior to cirrus subtraction, since this subtraction is not done in the SDSS data. The general picture shows a very good agreement between our data and SDSS data up to $R\sim300$ arcsec (i.e., equivalent to $\mu_{g,r}\sim27\,\rm{mag\, arcsec^{-2}}$), which roughly corresponds to the end of the inner disc. Beyond this radius, the surface brightness profiles of the SDSS data get systematically fainter than the profiles from our data, specially the $g$ band profile in sector A. The color profiles show more divergence between SDSS and our dataset. While our dataset, despite the presence of the Galactic Cirrus, present a $(g-r)_{0}\sim0.3$ plateau between 20 and 25 kpc associated with the blue disc presented in Sec.~\ref{sec:res}, the profiles from SDSS become redder beyond the limits of the inner disc. This shows how the reduction process plus the shallower data of SDSS presumably destroys the blue component observed in our dataset.
\begin{figure*}
    \sidecaption
     \includegraphics[width=12cm]{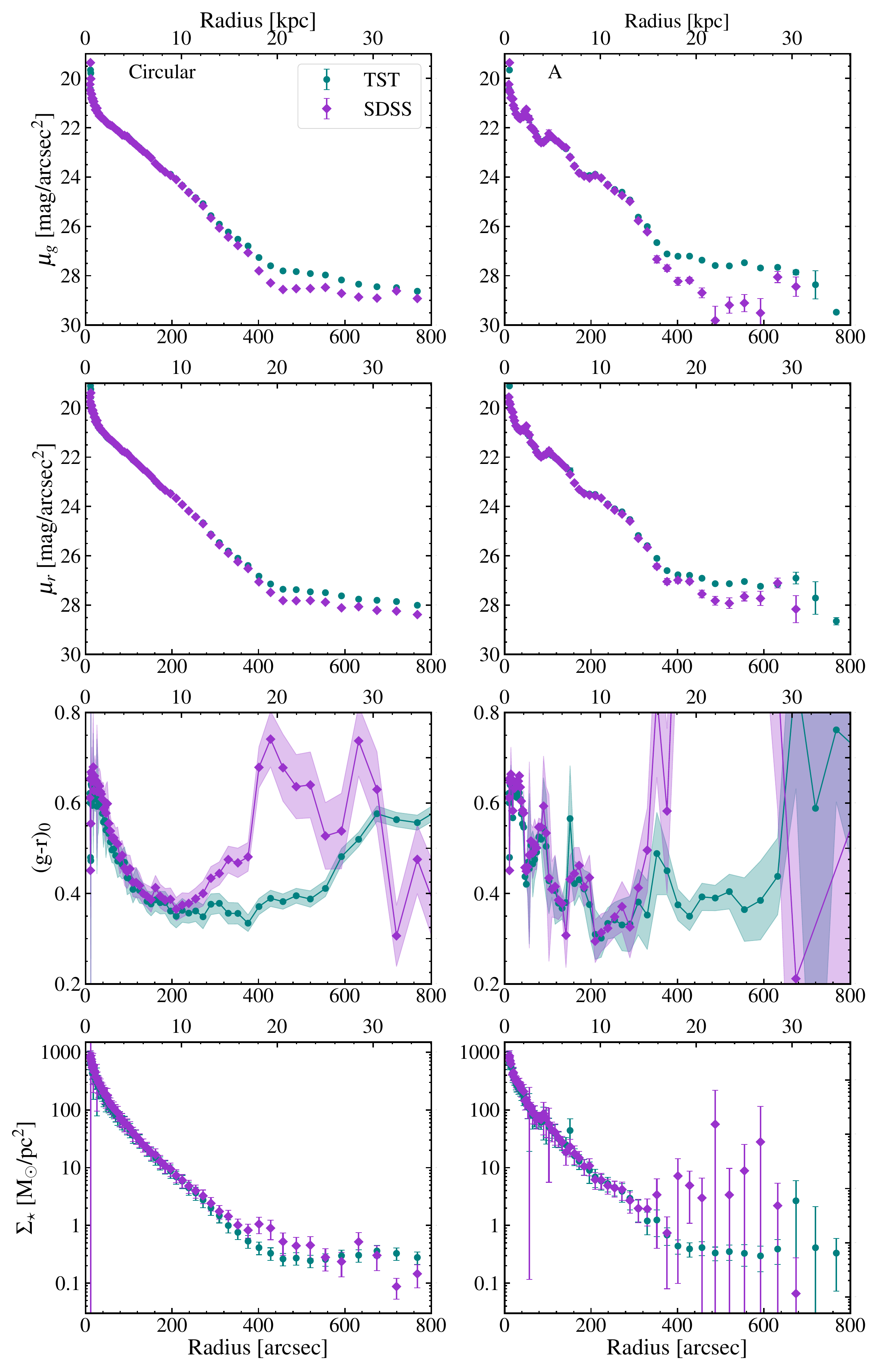}
     \caption{Surface brightness, colour and stellar surface density profiles of the TST data (teal dots) compared with those of the SDSS data (purple diamonds). The left panels correspond to a circular radial profile, while the right panels correspond to the profiles extracted from sector A.}
     \label{fig:sdssvststprofs}
\end{figure*}
\subsection{The size of M74 in \HI, UV and optical}\label{disc:size}
The extent of the M74 disc has been a matter of debate over the past few decades. \citet{kamphuis92} reported and analysed the large extent of the \HI disc (a maximum \HI radius of $60\,\rm{kpc}\sim3\times R_{Ho}$, being $R_{Ho}=6\arcmin$ the Holmberg radius of M74). They also noted the asymmetry of this disc, caused by a south-western tail, and the presence of symmetrically distributed high-velocity complexes (HVCs). They suggested this was proof that M74 is still accreting material from the intergalactic medium. Regarding the origin of these particularities in the \HI component of M74, they proposed three possible scenarios: a close encounter with the UGC~1171/1176 pair ($\sim$1 Gyr ago), a primordial origin or the accretion of a large cloud or an \HI-rich irregular dwarf galaxy.  \\

In the ultraviolet, \citet{thilker07} reported a Type I extended UV disc, showing that the stellar emission of this galaxy extends far beyond the reported size in optical images. However, the optical stellar emission had not yet been fully covered. Now, our ultra-deep optical data reconcile the galaxy's apparent size with the observed \HI and UV extents. At the position of the first break mentioned in Sec.~\ref{sec:res} ($R_{edge,in}=14\,\rm{kpc})$, the \HI density value is $\Sigma_{\HI}\sim5\,M_{\odot}\,\rm{pc}^{-2}$ in both the circular and the A sector profiles. At the end of the stellar density profile (i.e., at $R=30\,\rm{kpc}$), the \HI density reaches values of around $0.4\, M_{\odot}\,\rm{pc}^{-2}$. In Fig.~\ref{fig:hicontours}, we overplot the \HI density contours onto the optical image. The first contour, at $5\, M_{\odot}\,\rm{pc}^{-2}$, encloses the inner disc and the brightest regions of the blue spiral arms that extend beyond this inner disc. The second contour, at $\Sigma_{\HI}=0.4\, M_{\odot}\,\rm{pc}^{-2}$, covers both the inner disc and the new fainter regions observed in our optical data. Thus, these faint optical blue regions trace the previously known \HI extent of M74.\\

\begin{figure}
    \centering
    \includegraphics[width=\linewidth]{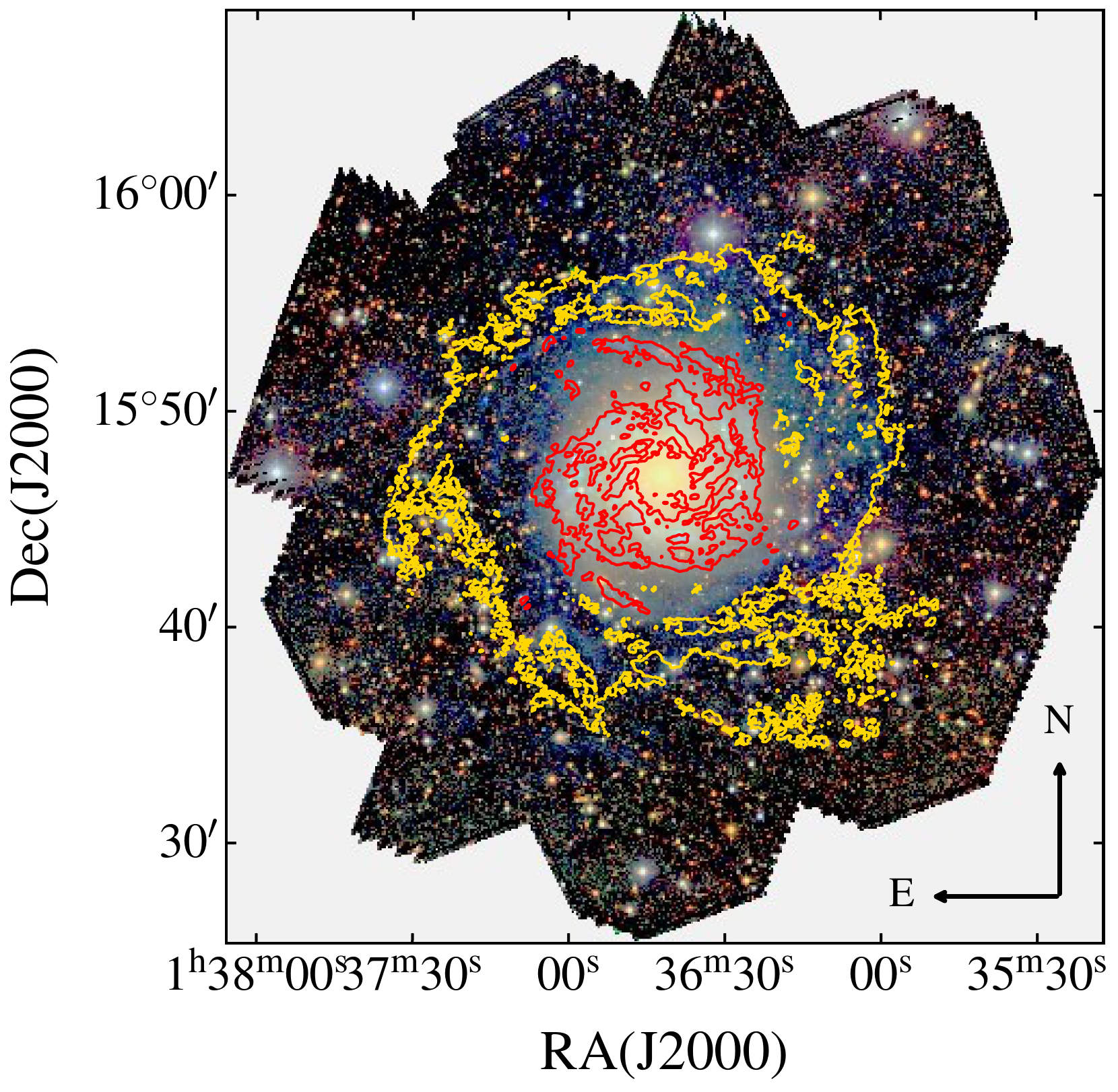}
    \caption{Contours of $\Sigma_{\rm{\HI}}=5\,\rm{M}_{\odot}\rm{pc}^{-2}$ (red) and $\Sigma_{\rm{\HI}}=0.4\,\rm{M}_{\odot}\rm{pc}^{-2}$ (yellow) overlaid on the optical deep colour image.}
    \label{fig:hicontours}
\end{figure}

Since the new M74 extension is way much larger than previous determinations based on shallower data, we explore where the new size estimation locate M74 in the mass-size relations (see Fig.~\ref{fig:masssize}). To compute the stellar mass at different radii, we integrated the circular apertures stellar surface density profile. This yielded masses of $M_{*}(R\leq14\,\mathrm{kpc} \equiv  R_{edge,in})\simeq1.25\times10^{10}\,M_{\odot}$, $M_{*}(R\leq28\,\mathrm{kpc}\equiv R_{edge,out})\simeq1.28\times10^{10}\,M_{\odot}$ and $M_{*}(R\leq14.4\,\mathrm{kpc}\equiv\,R_{1})\simeq1.25\times10^{10}\, M_{\odot}$. In Fig.~\ref{fig:radii}, we overplot these different radii as circles onto the image of M74 . \\

When using only the extension of the "old" disc, the size of the galaxy is located in the lower envelope of the mass-size relation, showing a trend similar to that of low T-type galaxies. However, when considering the extension of the young disc, M74 is placed in the upper envelope of the mass-size relation \citep{chamba22}. The new location of M74 in the upper envelope of the mass-size relation is suggestive of M74 being  caught during the formation of a very young and extended disc representative of the largest nowadays galaxies for its  stellar mass. At same stellar mass, bluer and younger stellar discs tend to present higher $R_{edge}$ in Fig. 5 of \citet{chamba22} (bottom left panels). The same trend is observed when comparing galaxies at different redshifts. In \citet[Fig. 5]{2024Buitrago}, galaxies at higher redhsifts populate the low size region of the mass-size relation. Thus, the expectation is that galaxies in the upper envelope of the mass-size relation are formed by large and young discs.   In this context, M74 would be one of the most recent newcomers to that region of the present-day mass-size plot. \\

\begin{figure*}
    \sidecaption
    \includegraphics[width=12cm]{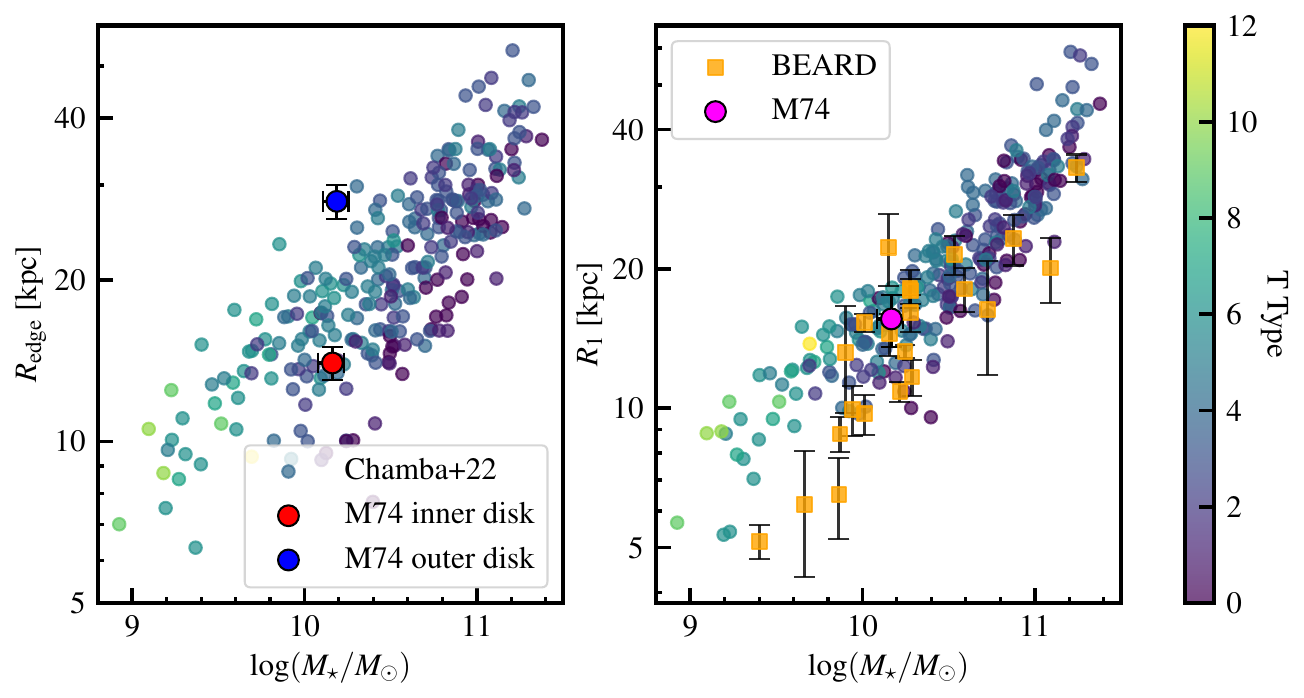}
    \caption{M74 in the different mass-size relations from \citet{2020trujillo,chamba22} for late-type galaxies (T>0). The left panel shows the location of the inner (i.e., old disc) and outer (i.e., young disc) edges defined in Sec.~\ref{sec:res} compared to $R_{edge}$ from the sample by \citet{chamba22} The right panel shows $R(\Sigma_{*}=1\,M_{\odot}\,\rm{pc}^{-2})$ for M74. We also plot BEARD galaxies \citep{carlos26}, given that M74 is reported to be bulgeless \citep{2020Herrera}.}
    \label{fig:masssize}
\end{figure*}
\begin{figure}
    \centering
    \includegraphics[width=\linewidth]{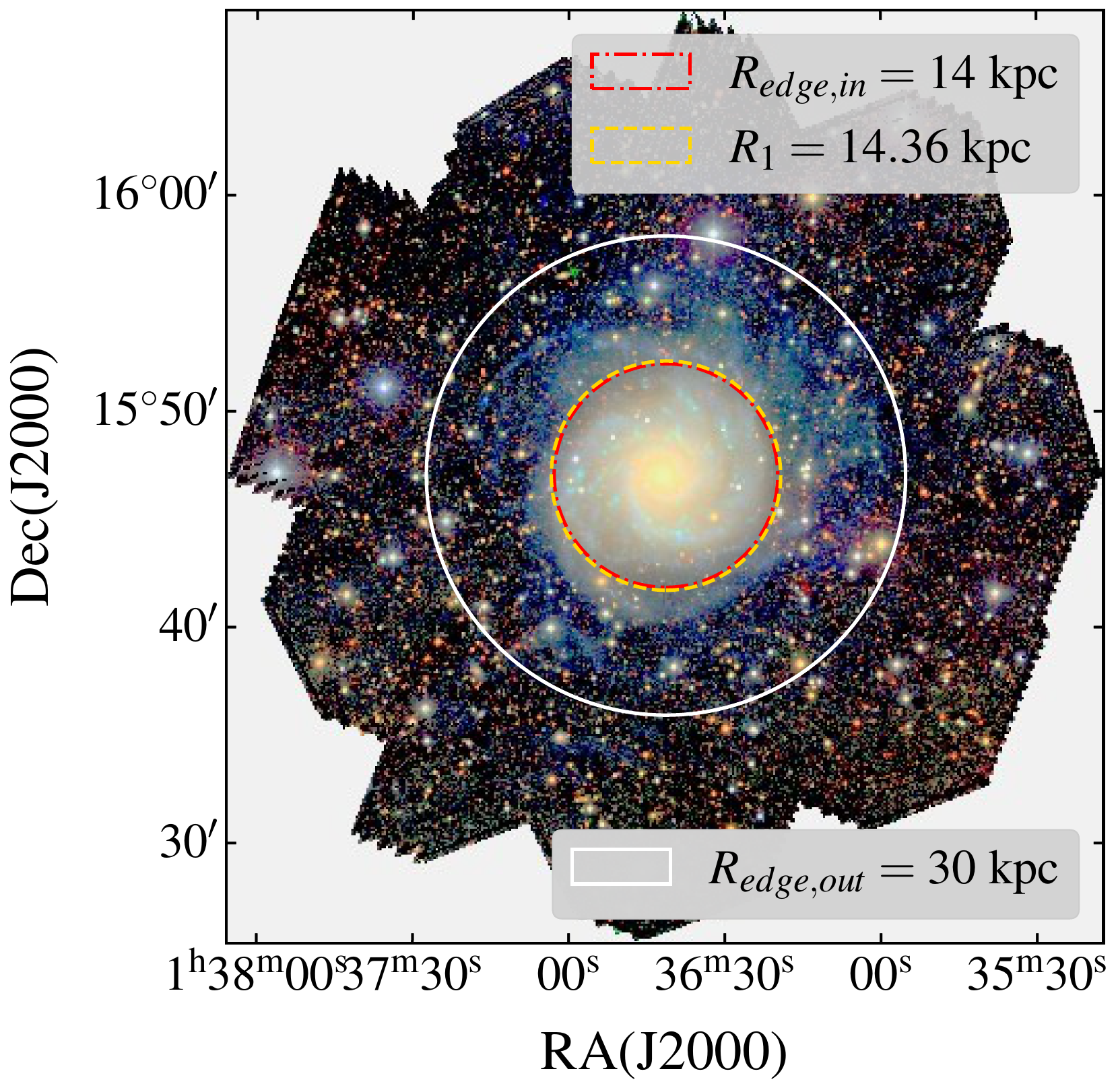}
    \caption{Circles representing the different radii defined for the mass-size relation overlaid on the colour image of M74: $R_{edge,\, in}=14$ kpc, $R_{1}=14.36$ kpc and $R_{edge,\,out}=30$ kpc. }
    \label{fig:radii}
\end{figure}
As a further support to our interpretation that M74 is a newcomer to the mass-size relations, the location of M74 in the $R_{1}$-mass relation follows the trend of the \citet{2020trujillo} sample. Interestingly, $R_{1}$ is close to the boundary of the inner old disc (see Fig.~\ref{fig:radii}). According to \citet{2020trujillo}, $R_{1}$ is a good proxy for the location where the gas density reaches the star formation threshold ($\Sigma_{*}=1\,M_{\odot}\,\rm{pc}^{-2}$ for late-type galaxies with an efficiency of $10-30\%$). \citet{2024Buitrago} showed that this threshold evolves from higher values at higher redshifts. In this work, M74 young disc shows star formation in the lower envelope of the reported threshold for local-Universe late-type galaxies \citep{chamba22}, with $\Sigma_{*}(R=R_{edge,out})\sim0.3\,M_{\odot}\,\rm{pc}^{-2}$. This might therefore represent a new generation of discs where the star formation density threshold is lower, following the evolution reported by \citet{2024Buitrago}. Surprisingly, this star formation is also below the \HI density threshold for $z=0$ \citep[$3$-$10\,M_{\odot}\,\rm{pc}^{-2}$, ][]{2004Schaye}. Overall, the emerging picture of the evolution of M74 is as follows: the old disc of the galaxy formed early on, with an extension (R$_{edge}$$\sim$14 kpc) and a stellar mass surface density at the edge $\Sigma_{*}(R=R_{edge,old})\sim1\,M_{\odot}\,\rm{pc}^{-2}$ representative of the conditions in the Universe around 1 Gyr ago \citep[see Figure 7 of ][]{2024Buitrago}.  Very recently, the disc extension has almost doubled in size, taking on the properties of the most recently formed galactic discs.\\

\subsection{Possible origin of the new disc}\label{disc:origin}

As mentioned in Sec.~\ref{disc:size}, \citet{kamphuis92} proposed three different scenarios to explain the large amount of \HI of M74: an interaction with satellites, a primordial origin, or accretion. \citet{2020Michalowski} also suggested an interaction (flyby) with UGC~1176 or NGC~660 as the most probable scenario, supported by the asymmetries in both the \HI and optical counterparts of M74, without ruling out the accretion scenario. We can explore these scenarios further by combining our deep optical data with the UV observations. \\

In Sec.~\ref{sec:res}, we showed that beyond $R\sim14$ kpc and up to $R\sim30$ kpc there is a new region of stellar emission in both optical and UV data that was not observed in previous surveys (Fig.~\ref{fig:sdssvstst}). We can combine our deep data with the UV data to compute a mean stellar age of this stellar population. We have measured the $(g-r)_{0}$ and $(FUV-NUV)_{0}$ colours of the north east spiral arm observed in our deep image (see Fig.~\ref{fig:m74}). The values obtained are $(g-r)_{0}=0.14\pm0.31$ and $(FUV-NUV)_{0}=0.2\pm1.5$. The high uncertainties of the measurements are connected to the low signal to noise of the outermost spiral arm. We then over-plot this point in a $g-r$ vs $FUV-NUV$ map where we also plot the evolutionary tracks computed from the \citet{bc03} models with low metallicites (see Fig.~\ref{fig:agecompute}). A $\chi^{2}$ test returns an age value of $\tau=640^{+375}_{-526}\,\rm{Myr}$ for a metallicity of $[\rm{Fe}/\rm{H}]=-1.65$. This value varies with metallicity, from $\tau=1015^{+263}_{-1012}\,\rm{Myr}$ for the lowest metallicity ($[\rm{Fe}/\rm{H}]=-2.25$); to $\tau=262^{+142}_{-106}\,\rm{Myr}$ for the highest metallicity ($[\rm{Fe}/\rm{H}]=1.00$). Despite this variation, all values are consistent with a $\tau\lesssim1\,\rm{Gyr}$ stellar population. We adopt the metallicity value of $[\rm{Fe}/\rm{H}]=-1.65$ as a reasonable assumption for a new stellar population with low metal enrichment.   \\

\begin{figure}
    \centering
    \includegraphics[width=\linewidth]{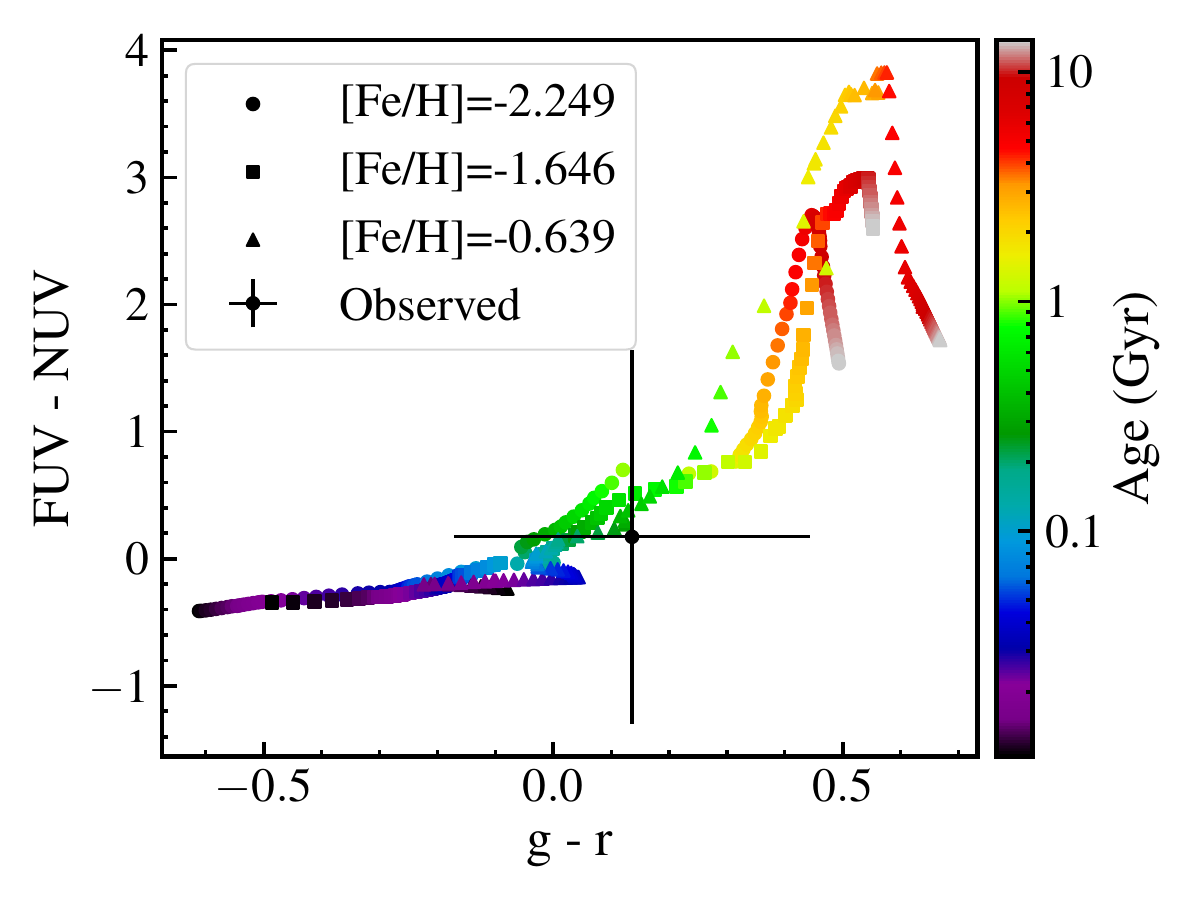}
    \caption{$g-r$ vs $FUV-NUV$ evolutionary tracks of the \citet{bc03} models for three different metallicities: $[\rm{Fe}/\rm{H}]=-2.25,-1.65$ and $-0.64$. The colour of the spiral arm is overplotted as a black dot.}
    \label{fig:agecompute}
\end{figure}

To explain the explosive growth of M74, the flyby scenario emerges as the most likely one, due to the asymmetries in both the \HI and optical discs. In the field, we identified three possible candidates: UGC~1171, UGC~1176 and SDSS~J013800.30+145858.1. All of these are situated at a projected distance of $\sim130\,\rm{kpc}$ from the centre of M74. Assuming a velocity dispersion for a Milky Way-like galaxy halo of $\sigma=136\,\rm{km\,s^{-1}}$ \citep{2004Hoekstra}, an object at such a distance could have interacted with M74 $\sim930\,\rm{Myr}$ ago. If the interaction triggered star formation in an extended quiet \HI disc around M74, then it would have propagated along the entire outskirts of the disc in $\sim530\,\rm{Myr}$  (assuming a circular radial velocity of $170\,\rm{km\,s^{-1}}$, \citealt{kamphuis92}, at a radial distance of 15 kpc). Based on these upper limit numbers, the age of the stellar population is consistent with a flyby scenario that has destabilized the \HI outer disc of M74. Of the objects mentioned before, we propose UGC~1176 as the most probable candidate. UGC~1171 and SDSS~J0138 are compact, round dwarf galaxies (see Fig.~\ref{fig:tstfield}) with no imprints of a turbulent past. On the contrary, UGC~1176 shows an irregular shape and bluer colours, compatible with a star formation also triggered by the interaction with M74. 

\section{Summary and Conclusions}\label{sec:conclusions}

In the recent years, the advent of deep optical imaging has enabled the study of the stellar populations at the outskirts of galaxies. These stellar populations are key to understanding the evolution of galactic discs. If formed \textit{in-situ}, they represent the latest imprint of the growth of the disc. In this work, we present the analysis of deep optical imaging of the galaxy Messier 74 (NGC~628). Previous works have reported a very extended and disturbed \HI disc \citep[see, e.g., ][]{kamphuis92,2006Auld}, as well as an extended UV disc \citep{thilker07}. However, no optical faint counterpart was detected in previous surveys such as SDSS. \\

Our new data show an extended ($\Delta R\sim16$ kpc), blue ($(g-r)_{0}\sim0.15$), very faint ($\mu_{g}\gtrsim27\,\rm{mag\,arcsec^{-2}}$)  region surrounding the known disc. Despite the contamination from Galactic Cirrus surrounding the galaxy, we propose that this new region is a recently ($\sim640\,\rm{Myr}$) formed extension of the stellar disc of M74. The colour, which is different from that of the Cirrus, and the shape, which is contained within the $\Sigma_{\HI}=0.4\,M_{\odot}\,\rm{pc}^{-2}$ contour, support this hypothesis. In addition, the optical extent is also compatible with the UV emission reported by \citet{thilker07}.  \\

The new extension of M74 is nearly a factor of two larger than its previously reported size. While both the edge radius of the  previously reported disc and $R_{1}$ are in good agreement with the mass-size relations of \citet{chamba22}, the new galaxy edge, at 30 kpc, places M74 in the upper envelope of the mass-size relation. The stellar mass surface density at the new edge is also in the lower envelopes of those reported by \citet{2020trujillo} and \citet{chamba22}, indicating that $R_{1}$ is not representative of the new disc. We propose that this reflects the evolutionary trend of disc galaxy edges over time, as reported by \citet{2024Buitrago}. We measured a mean age for this new disc of 640 Myr assuming a low-metallicity ($[\rm{Fe}/\rm{H}]=-1.65$), single stellar population. Given this and the visual inspection of the field, we propose a recent flyby of UGC~1176 at least 1 Gyr ago as the trigger for the new star formation in the outer regions. \\

In summary, in this work we propose an explosive evolution in the extension of M74. In less than 1 Gyr, a new disc of low-metallicity stars has formed, enlarging the galaxy up to $R\sim30\,\rm{kpc}$, which is twice as large as the old disc ending at $R\sim14\,\rm{kpc}$. Future studies will confirm whether this mechanism is present in other Local Universe galaxies, or if M74 is a unique case in galaxy evolution.

\section{Data availability}
The deep $g$, $r$ and $i$ images presented in Fig.~\ref{fig:tstfield} and the cirrus subtracted images of Sec.~\ref{sec:cirrus} are available at the CDS via anonymous ftp to \url{cdsarc.u-strasbg.fr} (130.79.128.5) or via \url{http://cdsweb.u-strasbg.fr/cgi-bin/qcat?J/A+A/}.

\begin{acknowledgements}
      We thank the referee for his detailed reading of the manuscript, which helped to clarify and improve the presentation of the results of this work. This article is based on observations made in the Transient Survey Telescope (TST) sited at the Teide Observatory of the Instituto de Astrofísica de Canarias (IAC), that Light Bridges operates in Tenerife, Canary Islands (Spain). The Observing Time Rights (DTO) used for this research were consumed in the PEI “GALAXDIF25". This research used storage and computing capacity in ASTRO POC’s EDGE computing centre at Tenerife under the form of Indefeasible Computer Rights (ICR). The ICR were consumed in the PEI "GALAXDIF25" with the collaboration of Hewlett Packard Enterprise and VAST DATA. Dr. Antonio Maudes' insights in economics and law were instrumental in shaping the development of this work. IRC and SGA acknowledge support from grant PID2022-140869NB-I00 from the Spanish Ministry of Science and Innovation. IT acknowledges support from the State Research Agency (AEI-MCINN) of the Spanish Ministry of Science and Innovation under the grant PID2022-140869NB-I00 and IAC project P/302302, financed by the Ministry of Science and Innovation, through the State Budget and by the Canary Islands Department of Economy, Knowledge, and Employment, through the Regional Budget of the Autonomous Community. This research also acknowledge support from the European Union through the following grants: "UNDARK" and "Excellence in Galaxies - Twinning the IAC" of the EU Horizon Europe Widening Actions  programmes (project numbers 101159929 and 101158446). Funding for this work/research was provided by the European Union (MSCA EDUCADO, GA 101119830). Views and opinions expressed are however those of the author(s) only and do not necessarily reflect those of the European Union or European Research Executive Agency (REA). Neither the European Union nor the granting authority can be held responsible for them.
\end{acknowledgements}

\bibliographystyle{aa}
\bibliography{bib.bib}

\begin{appendix}
\nolinenumbers
\onecolumn
\section{Comparison original image versus the one with the Cirrus removed}\label{ap:cirrus}
In Sec.~\ref{sec:cirrus}, we presented a technique outlined by \citet{kaboud26} to mititgate the effect of Galactic Cirrus. In Fig.~\ref{fig:ap_cirrus} we compare the profiles presented in Sec.~\ref{sec:res} with the ones obtained without applying this technique. Both $g$ and $r$ surface brightness profiles are affected by the Galactic Cirrus in its entirety, but the effect is mostly visible beyond $R\sim300\,\rm{arcsec}$.  This effect is important in the colour profiles, with differences up to $(g-r)_{0}\sim0.2-0.3$ in the outermost regions of M74, where the new disc is reported.

\begin{figure}
    \sidecaption
    \includegraphics[width=12cm]{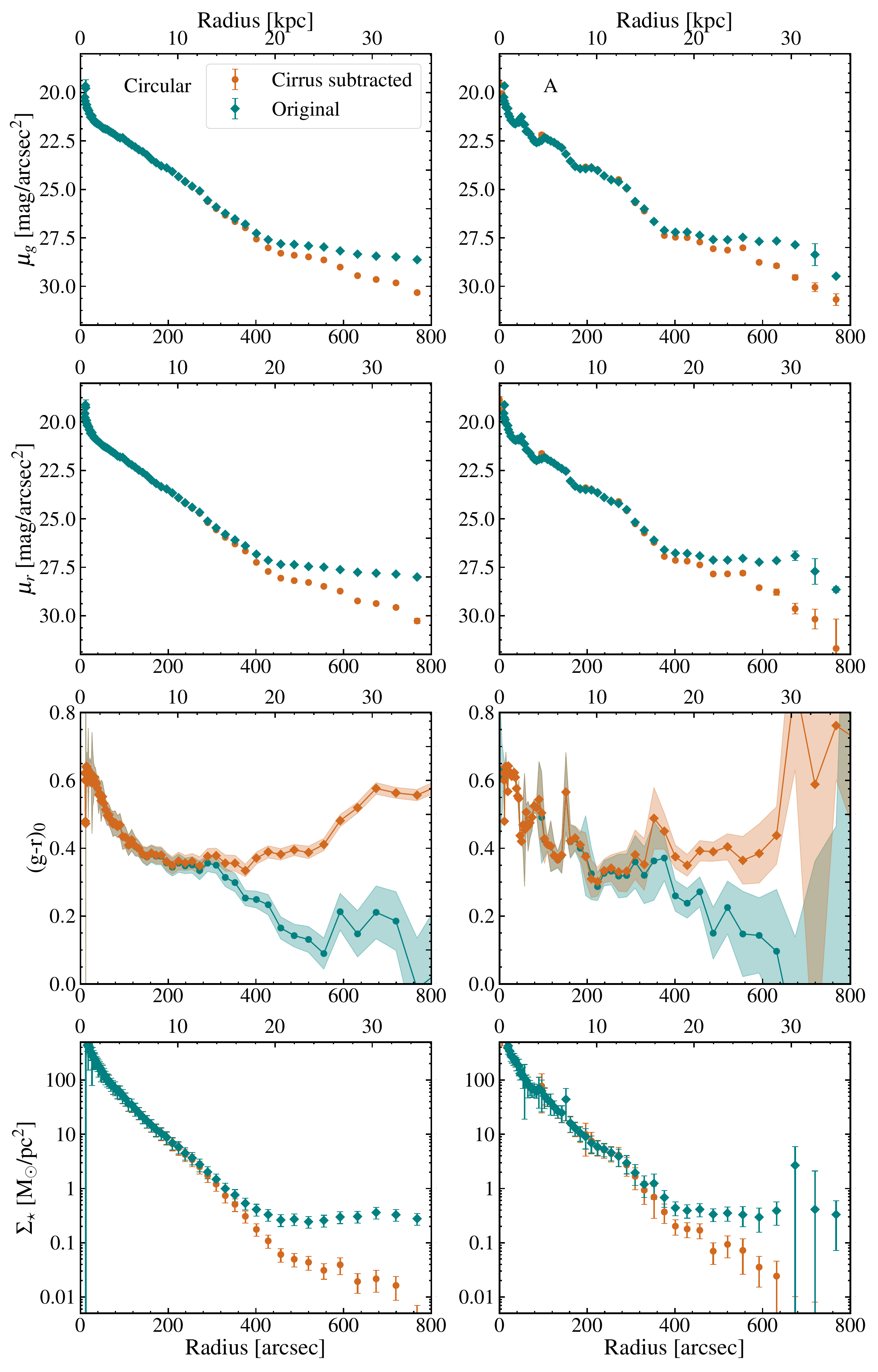}
    \caption{Surface brightness, colour and stellar mass density profiles of our optical data before (\textit{Original}, teal diamonds) and after (\textit{Cirrus subtracted}, orange dots) Cirrus subtraction. Left panels show the profiles using circular apertures, and right panels using the wedge-shaped sector A (see Fig.~\ref{fig:radprofs}).}
    \label{fig:ap_cirrus}
\end{figure}
\end{appendix}

\end{document}